\documentclass[aps,floatfix,preprint,preprintnumbers,nofootinbib,superscriptaddress,natbib]{revtex4}

\usepackage{graphicx,float}
\usepackage{epsfig}		
\usepackage{dcolumn}
\usepackage{bm}
\usepackage{subfigure}

\usepackage[all]{xy}
\usepackage{amsmath,upgreek}
\usepackage{amssymb}

\usepackage{pdfpages}
\usepackage{color}
\usepackage{graphicx,epstopdf}
\usepackage[colorlinks=true,
 linkcolor=red,
 urlcolor=purple,
 citecolor=blue,hyperindex]{hyperref}

\floatstyle{ruled}
\def\0{\mbox{\tiny $0$}}
\def\1{\mbox{\tiny $1$}}
\def\2{\mbox{\tiny $2$}}
\def\3{\mbox{\tiny $3$}}
\def\4{\mbox{\tiny $4$}}
\def\5{\mbox{\tiny $5$}}
\def\6{\mbox{\tiny $6$}}
\def\7{\mbox{\tiny $7$}}
\def\8{\mbox{\tiny $8$}}
\def\9{\mbox{\tiny $9$}}

\def\f14{\mbox{\tiny $\frac{1}{4}$}}

\begin{document}

\title{On dark sector scalar field theories driven by cosmological $c$-fields}

\author{Alex E. Bernardini}
\email{alexeb@ufscar.br}
\altaffiliation[On leave of absence from]{~Departamento de F\'{\i}sica, Universidade Federal de S\~ao Carlos, PO Box 676, 13565-905, S\~ao Carlos, SP, Brasil.}
\author{O. Bertolami}
\email{orfeu.bertolami@fc.up.pt}
\altaffiliation[Also at~]{Centro de F\'isica das Universidades do Minho e do Porto, Rua do Campo Alegre 687, 4169-007, Porto, Portugal.} 
\affiliation{Departamento de F\'isica e Astronomia, Faculdade de Ci\^{e}ncias da Universidade do Porto, Rua do Campo Alegre 687, 4169-007, Porto, Portugal.}

\date{\today}
\renewcommand{\baselinestretch}{1.3}

\begin{abstract}
The interplay of Hoyle-Narlikar (HN) creation field cosmology and scalar field models for the dark sector, including the generalized Chaplygin Gas (GCG), is investigated . Though originating from distinct theoretical frameworks, both the inclusion and the non-inclusion of the creation field degree of freedom (DoF) involve a scalar DoF, which addresses some of the limitations of the standard $\Lambda$CDM model. Using a Lagrangian scalar field formulation and the first-order Hamiltonian reconstruction method, the HN $c$-field dynamics is shown to be encompassed by the GCG equation of state through an equivalent modified scalar field theory. Late-time acceleration and stability of linear perturbations are derived within this unified description. Our results suggest that creation field cosmologies may be embedded in a broader class of scalar field models which encompasses subtle modifications to the Hubble expansion rate and related physical observables.
\end{abstract}

\date{\today}

\maketitle
\renewcommand{\baselinestretch}{1.7}

\section{Introduction}

The development of modern cosmology is deeply rooted in General Relativity, particularly in the solutions found by Friedmann and Lema{\^\i}tre, which describe a dynamic universe in expansion. When these solutions are combined with observational evidence --- especially the redshift-distance relation identified by Hubble --- they suggest a universe evolving from a highly dense and compact state. This reasoning underpins the so-called Big Bang model \cite{BB}. Despite its success in explaining features such as the cosmic microwave background (CMB) and the relative abundance of light elements, the standard cosmological model relies heavily on extrapolations beyond known physics, and presents conceptual challenges, such as the initial singularity, the flatness and horizon problems, and the unexplained nature of the dark sector.

An alternative line of investigation was proposed in the mid-20th century by Hoyle, Bondi, and Gold in the form of the Steady State theory \cite{Hoyle48,cct1}. Building on this, Hoyle and Narlikar introduced a theoretical framework based on different starting premises --- known as Creation Field Cosmology --- which result in slightly modifying Einstein's field equations to include a negative-energy scalar field ($c$-field) contribution responsible for continuous matter creation \cite{cct2,cct21,cct4}. This model avoids the initial singularity and offers solutions to some issues present in the standard Big Bang scenario. In particular, the continuous creation mechanism ensures a constant mean matter density despite the Universe's expansion --- a feature that contradicts observations.

Subsequent theoretical developments have expanded the scope of this so-called Hoyle-Narlikar (HN) framework \cite{cct2,cct21,cct4}. These include a Machian interpretation of inertia \cite{cct9}, conformal invariance \cite{cct11}, and compatibility with particle physics scales \cite{cct10}. Solutions involving negative-energy scalar radiation fields have been proposed as mechanisms for both matter creation and large-scale isotropy \cite{cct5,cct7}. Extensions to include higher-dimensional spacetimes \cite{cct18}, variable cosmological constants \cite{cct15,cct16,cct17}, and anisotropic geometries \cite{cct7} demonstrate the model's flexibility. The creation field formalism has also been applied to analyze gravitational wave backgrounds \cite{cct19} and baryon non-conservation \cite{cct20,cct21,cct22}.

Subtly, the HN theory also provides estimates for the CMB temperature that align closely with current measurements, even in the absence of a primordial Big-Bang scenario. For example, helium production in stars can yield a radiation background consistent with the observed blackbody temperature, suggesting that astrophysical processes alone may account for the CMB \cite{cct1}. Additionally, studies on the angular power spectrum within the quasi-steady state cosmology have also been considered for the explanation of CMB fluctuations \cite{Narlikar2003}. Otherwise, the criticism concerning the radio source counts was addressed in Ref.~\cite{qssc1994a}. 

The similarities between creation field cosmology and scalar field dark energy models --- particularly in their treatment of non-conventional energy components, dynamical evolution, and resolution of cosmological fine-tuning problems --- motivate a deeper investigation into possible theoretical connections. Both frameworks introduce scalar degrees of freedom to account for cosmic acceleration and structure formation, even though from distinct theoretical premises: the former from modifications to General Relativity and Machian principles, and the latter from effective field theory and high-energy physics.

In this work, a comparative analysis of the Hoyle-Narlikar creation field cosmology and scalar field cosmologies, including the generalized Chaplygin Gas (GCG) model \cite{Kam02,Bil02,bento2002}, is undertaken. The GCG model, governed by the equation of state $p \propto \rho^{-\alpha}$, with $0 < \alpha \leq 1$, suggests a unification of dark matter and dark energy into a single fluid, evolving from matter-like behavior at early times to a dark-energy-dominated regime at late times. This model also has been shown to be consistent with supernova, CMB, and other cosmological data within constrained parameter ranges \cite{bento2003,bento2003b,bertolami2006,bento2006}.
In this context, the $c$-field mathematical structure and its compatibility with cosmological predictions are examined. For this purpose, a Hamiltonian formulation that allows a generalized creation mechanism to be expressed in terms of the dynamics of scalar fields is considered. This approach aims to clarify whether the creation field concept can be embedded into broader scalar field cosmologies supported by General Relativity, suggesting the equivalence between these frameworks under a common theoretical structure.

The essential results of this letter are discussed at the end of Sec.~II, where the grounds for $c$-field cosmology are presented, their relations with the radiation dominated phase are identified, and the correspondence with scalar field cosmologies is demonstrated. In Sec.~III, by revisiting the first-order Hamiltonian method for reconstructing the field potential in terms of equivalent scalar fields, $\phi$ and $\varphi$, GCG cosmologies are shown to be capable of encompassing $c$-field contributions. The correspondence between the two equivalent scalar field theories is identified and the resulting contributions to the equation of state of the universe are computed. Fluctuations to the cosmological acceleration, $q$, and to the stability of the propagation of linear perturbations, as well as higher order effects described by jerk, $j$, and snap, $s$, parameters are then obtained and related to subtle modifications to the Hubble expansion rate, $H$.
Finally, our conclusions are drawn in Sec.~IV.

\section{$c$-field cosmology}

The theory of the quasi-steady state cosmology (QSSC) was developed in the early 1990s by Hoyle, Burbidge and Narlikar~\cite{qssc1993, qssc1994a, qssc1994b, qssc1995, qssc2000}. It is built upon the Machian theory of gravitation first proposed by Hoyle and Narlikar in the 1960s~\cite{cct2,cct21,cct4,hn1964, cct11}, where the origin of inertia is explained through a long-range scalar field interaction with matter.

The theory is based on two key principles. First, once causal propagation along light cones is conformal invariant and compatible with Lorentz invariance, the theory should be in line with electrodynamics, and must be conformally invariant. Specifically, if the spacetime metric changes from \( g_{\mu\nu} \) to \( \Omega^2 g_{\mu\nu} \), where \( \Omega \) is a non-vanishing function of spacetime coordinates, $X \equiv(x_0, x_1, x_2, x_3)$, the physical content of the theory should remain unchanged. This ensures that even if the local standard of length is altered from one spacetime point to another, the equations retain their form --- for example, in a theory with multiple frames related to multiple particles. 
Second, the theory should be Machian, that is, the inertia of a particle arises from the presence of all other particles in the universe. Thus, the world line of a particle \( a \) with inertial mass \( m_{(a)} \) at a given world-point \( A \) is subjected to contributions from all other particles \( b,\,c,\,\dots \). 

By defining the element of proper time of the particle $a$ along $A$ as $d\tau_a$, the theory can be derived from an action principle with the following form:
\begin{equation}\label{ap}
\mathcal{S} = -\sum_a \int m_{(a)} \, d\tau_a ,
\end{equation}
where the summation is over all particles in the universe labeled by $a$, each with mass $m_{(a)}$, and $d\tau_a$ is the element of proper time along the worldline of the $a$-th particle (i.e. \( b,\,c,\,\dots \)).

This is the full action for the theory and there is no separate gravitational term like in the Einstein-Hilbert action \cite{qssc1995}. The difference from Newtonian inertia is that the mass depends on the particle's position in spacetime, with its source being the distribution of all other particles.

Conformal invariance requires that the mass function scales by \( \Omega \), ensuring that, for instance, even the Dirac equation for a massive spinor remains conformally invariant. If one assumes that the Machian effect on each particle is mediated by a scalar wave equation with sources from other particles, the form of this wave equation is well-determined, as it will be identified below.

In this framework, the mass of a particle is not an intrinsic property, but arises from interactions with all other particles. Moreover, it is supposed that these particles are unstable and decay in an approximated Planckian scale, $\tau = 10^{-44} \,s$, producing $n = 6 \times 10^{19}$ baryons and radiation as secondary products of $a$, labeled by $a_1,\, a_2, \, \dots$. Specifically, the mass of particle $a$ at point $X$ on its worldline is given by:
\begin{equation}\label{mass}
m_{(a)}(X) = \sum_{r=1}^n m_{(a_r)}(X),
\end{equation}
where $m_{(a)}(X)$ is the contribution to the inertial mass from particle $a$ at a spacetime point $X$.
Such a Machian formulation intends to replace the principles of the Big-Bang cosmology and the Einstein-Hilbert derivation of the Einstein field equation. 

The typical particle to be created is short-lived and decays into more stable particles. The contributions to inertia of a typical particle come from all particles, stable or otherwise.
To implement such a dynamics, a conformally invariant scalar field $M(X) = c(X) + m(X)$ is introduced to describe the inertia at a spacetime point $X$, with $m(X) = \sum_a m_{(a)}(X)$. The contribution from $c(X)$ is due to short-lived primary particles, the so-called {\em Planck particles}, with squared mass $m^2_0 = 3h\mbox{c}/4\pi G$.

The mass function from Eq.~\eqref{mass} satisfies the conformally invariant scalar wave equation \cite{sachs1996}:
\begin{equation}\label{massss3}
\Box m_{(a)} + \frac{1}{6} R\, m_{(a)} + [m_{(a)}]^3 = N_{(a)},
\end{equation}
where $\Box$ is the d'Alembertian operator at point $X$, $R$ is the scalar curvature, and $N_{(a)}$ represents the number density of particle $a$,
\begin{equation}
 N_{(a)} =\sum_{r=1}^n\int_{A_o+\delta A_o}\frac{\delta_4(X,A_r))}{\sqrt{-g(A_r)}}d{\tau_{a_r}}.
\end{equation}

From Eq.~\eqref{massss3} one straightforwardly has the most general conformally invariant form for the wave equation for $m$,
\begin{equation}\label{massss}
\Box m + \frac{1}{6} R m + m^3 = \sum_a N_{(a)},
\end{equation}
where the so-called Planckian source spans along a time $\sim 10^{-44}\,s$ from world-point $A_o$ to $A_o+\delta A_o$ and the resulting decay products start at the world-point $A_o+\delta A_o$ and then continue onwards. 

The same argument of conformal invariance \cite{sachs1996,cct11}, can be considered for $c(X)$, 
\begin{equation}\label{file}
\Box c + \frac{1}{6} R c + c^3 = \sum_a \int_{A_o}^{A_o+\delta A_o}\frac{\delta_4(X,A))}{\sqrt{-g(A)}}d\tau_a,
\end{equation}
where it has been noticed that, if the Planck particles are labelled by $a$, $b$, $c$, etc., then one has $c(X) = \sum_a c_{(a)}(X)$.

The field equations being conformally invariant can be cast according to the choice of a specific conformal frame \cite{cct11} in which the part $m(X)$ is set constant. The contribution $c(X)$ is also negligible, but the related derivatives of the function, $c_{\mu}=\partial_{\mu} c$, are not negligible and they survive in the final field equation (cf. Sec.~VI of Ref.~\cite{cct11}),
\begin{equation}
\label{field}
R^{\mu\nu} - \frac{1}{2} g^{\mu\nu} R + \Lambda g^{\mu\nu} = - \kappa
\left[T^{\mu\nu} - \frac{2}{3} \left(c^{\mu} c^{\nu} - \frac{1}{4} g^{\mu\nu} c^\lambda c_\lambda \right)\right] \,,
\end{equation}
where $\kappa ={8 \pi G}$, $c$ is a scalar field associated with the creation (or annihilation) of particles, and light velocity and reduced Planck constant, $\mbox{c}$ and $\hbar$, have been set equal to unity, $\mbox{c}=\hbar=1$. The energy momentum tensor of all matter particles (the stable secondaries after the decay of unstable primaries) is captured by $T^{\mu\nu}$ \cite{sachs1996}.
Unlike in General Relativity, the energy-momentum tensor, $T^{\mu\nu}$, may have non-zero divergence, as it is compensated by the $c$-field's contribution. 
The Machian interpretation of Eq.~\eqref{field}, even if it is equivalently obtained by varying the action with respect to the metric $g_{\mu\nu}$, suggests that the constant of gravitation is related to the basic property of inertia, with the cosmological constant emerging from the cubic terms of all the wave Eqs.~\eqref{massss} and \eqref{file} put together.
The gravitational constant $G$ and cosmological constant $\Lambda$ are therefore derived quantities given by \cite{sachs1996} $G = {3}/{4 \pi m_0^2}$ and $\Lambda = -3/{\mathcal{N}^2 m_0^2}$, where $\mathcal{N}$ is the number of particles within the cosmic horizon, i.e. the primary Planck particles from which the secondaries are present in the observable universe, and the signs of these constants follow naturally from the theory: $G$ is positive, $\Lambda$ is negative, and the coupling of the $c$-field to the geometry is negative.

Despite its intricated inception, the above Machian motivated formulation arises from the equation of motion, Eq.~\eqref{field}, which can be confronted with the Einstein-Hilbert action approach of General Relativity as we will show in what follows.
 
\subsection{Quasi-steady state cosmology}

It follows from the action principle, Eq.~\eqref{ap}, that matter creation takes place at spacetime points where the ambient $c$-field reaches $c = m_0$. While the $c$-field typically remains below this threshold, strong gravitational fields near compact massive objects can locally enhance it, triggering the creation of matter and associated negative $c$-field energy. The negative pressure of this field drives an expansion analogous to inflation, ejecting matter outward in what is termed a ``minibang'' \cite{cct7,Hoyle1995,Narlikar2003}.

Spherically symmetric mass distributions (Schwarzschild-type) produce isotropic explosions, while axially symmetric ones (Kerr-type) yield jets along the axis of symmetry. Given angular momentum conservation during collapse, jet-like ejections are generally expected to dominate.

Importantly, these minibangs are non-singular. There is no infinite curvature or terminating worldlines as in the standard Big Bang, and no event horizon forms due to the repulsive effect of the $c$-field. The collapsing object undergoes a bounce before an event horizon can form.

On cosmic scales, these localized creation events drive the expansion of the universe. In the absence of such activity, spacetime tends toward a de Sitter-like expansion. However, creation events occur episodically, inducing oscillations around a long-term steady state. Sachs et al. \cite{sachs1996} found a simple solution, in such a context, for the FRW universe with line element of the geometry yielded as 
\begin{equation}
\label{plane}
ds^{2} = - dt^{2} + a(t)^2\,\left[dr^2 + r^2\,\left(d\theta^2 + \sin^2\theta d\phi^2 \right)\right].
\end{equation}
In this case, for an isotropic $c$-field, with components identified by $c_\mu= \delta_{0\mu}\dot{c}$, where $\dot{c}$ is the time-derivative of a scalar $c$, the scale factor $a(t)$ results in
\begin{equation}
a(t) = e^{t/P} \left[1 + n \cos\left( \frac{2\pi \sigma(t)}{Q} \right) \right],
\end{equation}
where constants $P$ and $Q$ relate to the fundamental parameters of the field equations, $\sigma(t)$ is a slowly varying function (approximately linear in $t$), and $n$ is a positive constant less than unity \cite{sachs1996}. The scale factor never vanishes, and the cosmological solution is entirely free of spacetime singularities.

\subsection{Radiation-dominated cosmology}

For the metric Eq.~\eqref{plane}, now with $\Lambda$ replaced by $\lambda^2$, i.e. a positive cosmological constant value, the field Eq.~\eqref{field} can be cast in terms of its time and space components as
\begin{eqnarray}
\label{Fri}
3\left(\frac{\dot{a}^2}{a^2} + \frac{k}{a^2}\right) &=& 3 \lambda^2 +\kappa \rho - \frac{2}{3}\frac{3\kappa}{4}\dot{c}^2,\\
\label{Ein}
2\frac{\ddot{a}}{a}+\frac{\dot{a}^2}{a^2} + \frac{k}{a^2} &=& 3 \lambda^2 -\kappa p + \frac{2}{3}\frac{\kappa}{4}\dot{c}^2,
\end{eqnarray}
where the background fluid energy density, $\rho\equiv\rho(a)$, and pressure, $p\equiv p(a)$, can be identified in terms of underlying fields.
Since in the universe dominated by the radiation phase one has $\rho = 3p$, Eqs.~\eqref{Fri}-\eqref{Ein} can be manipulated to give
\begin{equation}
\label{Rad}
\frac{\ddot{a}}{a}+\frac{\dot{a}^2}{a^2} + \frac{k}{a^2} - 2\lambda^2 = 0,
\end{equation}
from which one has, independently of $c(t)$, solutions given by
\begin{equation}
\label{scalarRad}
a(t) = \frac{e^{\lambda \left(t-t_0\right)} \sqrt{\left[e^{ 2 \lambda \left(t-t_0\right)}+k\right]^2- (1+k)^2+4\lambda^2\,a^2_0}}{2\lambda},
\end{equation}
with $\lambda >0$ and $a(t_0)=a_0$, which can be straightforwardly manipulated to give well-known solutions as, for instance, for a flat universe ($k=0$),
\begin{equation}
\label{scalarRadplane}
a(t) = a_0 \sqrt{\cosh[2\lambda(t-t_0)]},
\end{equation}
and for a strictly (open) radiation dominated universe ($\lambda \ll a^{-4}$, with $\lambda \to 0$),
\begin{eqnarray}
\label{scalarRad2A}
a(t) &=& a_0 \sqrt{t-t_0},\quad\mbox{for}\quad k = 0;\\
\label{scalarRad2B}
a(t) &=& \sqrt{(t-t_0)^2+a_0^2},\quad\mbox{for}\quad k = -1.
\end{eqnarray}

Of course, solutions for $c(t)$ can be straightforwardly obtained from Eq.~\eqref{Fri} once one has $\rho \equiv \rho(a)$. For a radiation dominated flat universe, from Eq.~\eqref{scalarRad2A}, one obtains $c(t) \propto \ln(t)$.

\subsection{Dark sector scalar field cosmology}

If the background fluid can be associated to a real scalar field, then the energy density, $\rho$, and pressure, $p$, can be written as
\begin{eqnarray}
\label{lcdm005B}
3\lambda^2/\kappa+\rho\equiv\rho_{\phi} &=& \frac{\dot{\phi}^{2}}{2} + V(\phi), \\
\label{lcdm005}
-3\lambda^2/\kappa+p\equiv p_{\phi} &=& \frac{\dot{\phi}^{2}}{2} - V(\phi),
\end{eqnarray}
and Eqs.~\eqref{Fri} and \eqref{Ein} are written as
\begin{eqnarray}
\label{FriPhi}
3\left(\frac{\dot{a}^2}{a^2} + \frac{k}{a^2}\right) &=& \kappa \left[\frac{\dot{\phi}^{2}}{2} + V(\phi) - \frac{1}{2}\dot{c}^2\right],\\
\label{EinPhi}
2\frac{\ddot{a}}{a}+\frac{\dot{a}^2}{a^2} + \frac{k}{a^2} &=& -\kappa \left[\frac{\dot{\phi}^{2}}{2} - V(\phi) - \frac{1}{6}\dot{c}^2\right],
\end{eqnarray}
with the cosmological constant contribution, $3\lambda^2/\kappa$, absorbed by the scalar field potential, $V(\phi)$, such that \footnote{One divides \eqref{FriPhi}-\eqref{EinPhi} by $d\eqref{FriPhi}/dt$.}
\begin{equation}
\ddot{\phi} + 3 H \dot{\phi} + \frac{d V(\phi)}{d {\phi}} = \frac{d c}{d {\phi}}\left(\ddot{c}+ 2 H \,\dot{c}\right),
\label{fluidphi}
\end{equation}
with $H=\dot{a}/a$, and from which, for vanishing $c$-field time derivatives, one recovers back the the perfect fluid energy-momentum conservation equation,
\begin{equation}
\dot{\rho_\phi} + 3 H (\rho_\phi + p_\phi) = 0.
\label{fluid}
\end{equation}
Interestingly, from Eq.~\eqref{fluidphi}, a corresponding ``twin'' scalar field theory can be obtained and recast in the form of Eqs.~\eqref{FriPhi} and \eqref{EinPhi}.
This can be yield through coupled constraint equations, 
\begin{eqnarray}
\label{kin}
\dot{\varphi}^{2} &=& \dot{\phi}^{2} -\frac{2}{3}\dot{c}^{2},\\
\label{pot}
U(\varphi) &=& V(\phi) - \frac{1}{6}\dot{c}^{2},
\end{eqnarray}
such that, from straightforward manipulations of Eq.~\eqref{fluidphi}, one obtains
\begin{equation}
\ddot{\varphi} + 3 H \dot{\varphi} + \frac{d U(\varphi)}{d {\varphi}} = 0,
\label{fluidvarphi}
\end{equation}
and therefore
\begin{eqnarray}
\label{FrivPhi}
3\left(\frac{\dot{a}^2}{a^2} + \frac{k}{a^2}\right) &=& \kappa \left[\frac{\dot{\varphi}^{2}}{2} + U(\varphi)\right],\\
\label{EinvPhi}
2\frac{\ddot{a}}{a}+\frac{\dot{a}^2}{a^2} + \frac{k}{a^2} &=& -\kappa \left[\frac{\dot{\varphi}^{2}}{2} - U(\varphi)\right],
\end{eqnarray}
from which one notices that the $c$-field degree of freedom is completely absorbed by the $\varphi$-scalar field theory.

\section{GCG and $c$-field cosmology}

Initially considering a $c$-field null contribution scenario, a typical self-contained $\phi$-scalar field theory describing a dark sector Universe \cite{Kam02,bento2002} can be evaluated by assuming that energy density and pressure are related through the generalized Chaplygin gas (GCG) equation of state,
\begin{equation}
p = - \Omega_{\Lambda} \rho^{0} \left(\frac{\rho^{0}}{\rho}\right)^{\alpha},
\label{lcdm001}
\end{equation}
for which an energy density which interpolates between a dust dominated phase, $\rho\propto a^{-3}$, in the past, and a de-Sitter phase, $\rho = -p = \Omega_{\Lambda}$, at late times \footnote{The constants $\Omega_{\Lambda}$ and $\alpha$ are positive and $0 < \alpha \leq 1$.
Of course, $\alpha = 0$ corresponds to the $\Lambda$CDM model and $\alpha = 1$ corresponds to the equation of state of the Chaplygin gas scenario \cite{Kam02,bento2002,Bil02}.}.
In fact, after inserting Eq.~\eqref{lcdm001} into the unperturbed energy conservation Eq.~(\ref{fluid}), through a straightforward integration, one obtains \cite{Kam02,bento2002}
\begin{equation}
\rho \equiv \rho_{\phi}(a)= \rho_{0} \left[\Omega_{\Lambda} + \frac{(1-\Omega_{\Lambda})}{a^{3(1+\alpha)}}\right]^{1/(1 + \alpha)},
\label{gcg21}
\end{equation}
and
\begin{equation}
p \equiv p_{\phi}(a)= - \Omega_{\Lambda} \rho_{0} \left[\Omega_{\Lambda} + \frac{(1-\Omega_{\Lambda})}{a^{3(1+\alpha)}}\right]^{-\alpha/(1 + \alpha)},
\label{gcg22}
\end{equation}
which makes the GCG model an interesting candidate for the unification of dark matter and dark energy through an underlying scalar field theory with specific features.

\subsection{First-order Hamiltonian formulation: scalar field cosmology} 

For spatially flat Friedmann-Robertson-Walker (FRW) cosmologies, the Hamilton-Jacobi framework offers a simplified approach for describing the dynamics of scalar field-driven expansion \cite{Ber2013}. This formulation rewrites the background cosmological evolution in terms of the dynamical scalar field $\phi(t)$ and the scale factor $a(t)$, both time-dependent quantities.
The GCG model can be recovered from a canonical scalar field model derived from a generalized Born-Infeld-type action \cite{bento2002,Bil02}, such that by substituting Eqs.~\eqref{lcdm005B} and \eqref{lcdm005} into the Eq.~\eqref{fluid}, one has
\begin{equation}
\ddot{\phi} + 3 H \dot{\phi} + \frac{d V(\phi)}{d {\phi}} = 0,
\label{fluidphiB}
\end{equation}
from which one obtains an implicit analytical solution which, through Eqs.~\eqref{gcg21} and \eqref{gcg22}, can be recast in terms of the scale factor, $a$, as
\begin{equation}
\phi(a) = \frac{1}{\sqrt{6} (\alpha + 1)}\ln{\left[\frac{\sqrt{1 - \Omega_{\Lambda}(1 - a^{3(\alpha + 1)})} + \sqrt{1 - \Omega_{\Lambda}}}{\sqrt{1 - \Omega_{\Lambda}(1 - a^{3(\alpha + 1)})} - \sqrt{1 - \Omega_{\Lambda}}}\right]},
\label{lcdm24}
\end{equation}
with the present-time value (at $a=a_0=1$) given by
\begin{equation}
\phi_{0} = \phi(1) = \frac{1}{\sqrt{6} (\alpha + 1)}\ln{\left[\frac{1 + \sqrt{1 - \Omega_{\Lambda}}}{1 - \sqrt{1 - \Omega_{\Lambda}}}\right]}.
\label{lcdm25}
\end{equation}

From this expression, the scalar potential $V(\phi)$ can be directly derived \cite{bento2002}:
\begin{equation}
V(\phi) =
\frac{1}{2}\Omega_{\Lambda}^{\frac{1}{\alpha + 1}}\rho^{0}
\left\{
\left[\cosh{\left(\sqrt{\frac{3}{2}}(\alpha + 1) \phi\right)}\right]^{\frac{2}{\alpha + 1}}
+
\left[\cosh{\left(\sqrt{\frac{3}{2}} (\alpha + 1) \phi\right)}\right]^{-\frac{2\alpha}{\alpha + 1}}
\right\}.
\label{lcdm26}
\end{equation}

Within the Hamilton-Jacobi formalism, the scalar field serves as the fundamental dynamical variable, while the Hubble expansion rate, $H$, is rewritten as a function of $\phi$:
\begin{equation}
H = \frac{\dot{a}}{a}= y(\phi),
\label{xx}
\end{equation}
and the evolution of $\phi$ is governed by the relation:
\begin{equation}
\dot{\phi} = \frac{d y}{d\phi} \equiv y_{\phi},
\label{xxx}
\end{equation}
leading to a set of first-order differential equations for the background evolution, where, from this point, $\kappa/2$ (cf. Eqs.~\eqref{FriPhi} and \eqref{EinPhi}) has been set equal to unity, $\kappa/2 = 4 \pi G \equiv 1$, as to keep $\rho_0$ given as a multiple of $\rho_{Crit} = 3H_0^2/8\pi G$ (with $\rho_0 = 1$ for $k =0$, in a flat universe) and $H$ as a multiple of the Hubble constant, $H_0$.

By combining Eqs.~\eqref{xx} and \eqref{xxx} with the scalar field dynamics described by Eq.~(\ref{FriPhi}), one obtains the potential as
\begin{equation}
V(\phi) = \frac{3}{2} y^{2} - \frac{1}{2} y_{\phi}^{2}.
\label{teste}
\end{equation}

Returning to the GCG model, a straightforward computation yields the functional form of $y(\phi)$ as
\begin{equation}
y(\phi) =
\left(\Omega_{\Lambda}^{\frac{1}{\alpha + 1}}\rho^{0}\right)^{\frac{1}{2}}
\cosh{\left[\sqrt{\frac{3}{2}} (\alpha + 1) \phi \right]}^{\frac{1}{\alpha + 1}},
\label{teste1}
\end{equation}
which, when substituted back into Eq.~(\ref{lcdm26}), reproduces the expected expression for $V(\phi)$.

\subsection{First-order Hamiltonian formulation with $c$-field cosmology} 

The generic first-order Hamiltonian formulation arising from Eqs.~\eqref{xx}-\eqref{teste2} results in the same scalar field dynamics described by Eq.~\eqref{fluidphi} if $\dot{c}$ had been turned off, i.e. $\dot{c} \sim 0$ \footnote{Or even if $$\ddot{c}+2 H \dot{c} = 0,$$ may correspond to an exponentially decaying $c$-field.}.
Turning on the $c$-field, one should replace $\phi$ by $\varphi$, and also that $H \equiv H_{(\phi)} = y(\phi)$ (cf. Eq.~\eqref{xx}) is replaced by
\begin{equation}
H_{(\varphi)}= w(\varphi),
\label{xx2}
\end{equation}
with the evolution of $\varphi$ driven by the relation
\begin{equation}
\dot{\varphi} = \frac{d w}{d\varphi} \equiv w_{\varphi},
\label{xxx2}
\end{equation}
which can be combined with the scalar field dynamics described by Eq.~(\ref{FrivPhi}), in order to give
\begin{equation}
U(\varphi) = \frac{3}{2} w^{2} - \frac{1}{2} w_{\varphi}^{2}.
\label{teste2}
\end{equation}

One should notice that $H_{(\phi)}\neq H_{(\varphi)}$, as a result of different time evolutions for $a(t)$. The exact correspondence between $\phi$ and $\varphi$ scalar field cosmologies described by the respective Hubble expansion rates, $H_{(\phi)}$ and $H_{(\varphi)}$, is just recovered through the inclusion of an intermediating $c$-field, which corresponds to an additional degree of freedom in the $\phi$ theory. In other words, this additional degree of freedom can be absorbed by the constraint between $\phi$ and $\varphi$ (cf. Eqs.~\eqref{kin} and \eqref{pot}), such that $\varphi \equiv \varphi(\phi)$.
In this case, one has 
\begin{equation}
\dot{\varphi} = \varphi_{\phi} \dot{\phi},\qquad \mbox{with} \quad \varphi_{\phi} \equiv \frac{d \varphi}{d\phi},
\label{ee1}
\end{equation}
and, from Eqs.~\eqref{xxx} and \eqref{xxx2},
\begin{equation}
w_{\varphi} = \varphi_{\phi} y_{\phi}
\label{ee2}.
\end{equation}
From Eqs.~\eqref{xx2}-\eqref{teste2} and Eqs.~\eqref{xx}-\eqref{teste}, once replaced into Eqs.~\eqref{kin} and \eqref{pot}, after some straightforward mathematical manipulations, one obtains
\begin{equation}
\dot{c}^2 = \frac{3}{2}y^2_{\phi}\left(1 - \varphi_{\phi}^2\right) = -\frac{3}{2}w^2_{\varphi}\left(1 - \phi_{\varphi}^2\right)
\label{ee1m}
\end{equation}
and
\begin{equation}
w^{2} = y^{2} - \frac{1}{2} y^2_{\phi}\left(1 - \varphi_{\phi}^2\right) \leftrightarrow y^{2} = w^{2} - \frac{1}{2} w^2_{\varphi}\left(1 - \phi_{\varphi}^2\right),
\label{ee2m}
\end{equation}
where the correspondence between $\varphi_{\phi} = \phi_{\varphi}^{-1}$ (or even $\varphi(\phi) \leftrightarrow \phi(\varphi)$) and $c$ constrains the solutions of the system of equations for both $\phi$ and $\varphi$ scalar field theories.

A suitable aspect to notice is the relation between $H_{(\phi)}$ and $H_{(\varphi)}$ obtained from Eqs.~\eqref{ee1m} and \eqref{ee2m} as
\begin{equation}
H^2_{(\varphi)} = H^2_{(\phi)} - \frac{\dot{c}^2}{3},
\label{ee2mBB}
\end{equation}
which shows how the $c$-field contributions can modify the Hubble expansion rate in the context of a scalar field theory for $\phi$, when mapped by another scalar field theory for $\varphi$ \footnote{Notice that $H^2_{(\varphi)}$ corresponds to the same $H$ that appears in Eqs.~\eqref{fluidphi} and \eqref{fluidvarphi}, which simplifies to $H^2_{(\phi)}$ for $\dot{c} =0$.}.

\subsection{Re-scaled and deformed scalar fields and the cosmological outputs for the GCG}

Assuming that two equivalent dark sector field theories can be described in terms of independent scalar fields, $\phi$ and $\varphi$, both connected by a $c$-field through the dynamical map of Eqs.~\eqref{ee1m} and \eqref{ee2m}, one can discuss the impact of a $c$-field onto the GCG scalar field theory.
This description consists in replacing the $c$-field degree of freedom by a correspondence between $\phi$ and $\varphi$ (i.e. by identifying $\varphi(\phi) \leftrightarrow \phi(\varphi)$ into Eq.~\eqref{ee1m}).

Moreover, from Eq.~\eqref{ee2mBB}, one obtains
\begin{equation}
H^2_{(\varphi)} = H^2_{(\phi)} - \frac{1}{2}y^2_{\phi}\left(1 - \varphi_{\phi}^2\right),
\label{ee2mcc}
\end{equation}
that is, the dark sector driven by $\phi$ is subtly modified by the choice of $\varphi(\phi) \leftrightarrow \phi(\varphi)$.

Considering GCG solutions, and solving the system of Eqs.~\eqref{ee1m} and \eqref{ee2m}, modifications on cosmological fluid can be examined by expanding the pressure (given the equation of state) in a Taylor series around the current energy density:
\begin{equation}
 p = p_0 + 
 \sum_{n=1}^{N-1} {1\over n!} 
 \left.{d^n p\over d\rho^n}\right|_0 (\rho-\rho_0)^n 
 + O[(\rho-\rho_0)^N],
\end{equation}
such that the $n$-th order Taylor coefficient,
\begin{equation}
\left.{d^n p\over d\rho^n}\right|_0,
\end{equation}
can be shown to depend on the ($n$+2)-th time derivative of the scale
factor
\begin{equation}
\left.{d^{n+2} a(t)\over d t^{n+2}}\right|_0,
\end{equation}
and hence to the ($n$+2)-th term, the $O(z^{n+2})$ term, in the
Taylor expansion of the Hubble expansion rate.

Drawing from classical mechanics, where successive time derivatives of position are referred to as velocity, acceleration, jerk, and snap, analogous cosmological quantities \cite{Visser} can be defined in terms of the scale factor:
\begin{eqnarray}
H(t) = + {1\over a} \; {d a\over d t};\qquad\qquad\,\,\,&&\\
q(t) = - {1\over a} \; {d^2 a\over d t^2}
\left[ {1\over a} \; {d a \over d t}\right]^{-2} &\equiv& q(a)= -\left[1+\frac{1}{2}\frac{d\ln(\rho(a))}{d\ln(a)}\right];
\\
j(t) = + {1\over a} \; {d^3 a \over d t^3} 
\; \left[ {1\over a} \; {d a \over d t}\right]^{-3}&\equiv& j(a)= -q(a)\left[1+\frac{d\ln(q(a)\rho(a))}{d\ln(a)}\right];
\\
s(t) = + {1\over a} \; {d^4 a \over d t^4} 
\; \left[ {1\over a} \; {d a \over d t}\right]^{-4}&\equiv& s(a)= j(a)\left[1+\frac{3}{2}\frac{d\ln(j(a)\rho(a))}{d\ln(a)}\right],
\end{eqnarray}
such that one can write
\begin{eqnarray}
a(t)= a_0 \;
\Bigg\{ 1 + H_0 \; (t-t_0) - {1\over2} \; q_0 \; H_0^2 \;(t-t_0)^2 
+{1\over3!}\; j_0\; H_0^3 \;(t-t_0)^3 
\nonumber
\\
\qquad
+{1\over4!}\; s_0\; H_0^4 \;(t-t_0)^4
+ O([t-t_0]^5) \Bigg\}.
\end{eqnarray}

\subsubsection{Results for a re-scaled scalar field: $\varphi_\phi = \ell$.}

The simplest constraint which can be considered is given by a re-scaled scalar field,
\begin{equation}
\varphi(\phi) = \ell\phi, 
\label{cfcvA3}
\end{equation}
where $\ell$ is an arbitrary constant with $0 < \ell < 1$.

The energy density dependence on the scale factor, $\rho_{\varphi}(a)$, can be straightforwardly identified from Eqs.~\eqref{xx2} and \eqref{ee2m}, as
\begin{eqnarray}
\rho_{\varphi}(a)
&=&\frac{3}{2}w^2(a)\nonumber\\
&=&\frac{3}{2}y^{2}(a) - \frac{3}{4} y^2_{\phi}(a)\left(1 - \varphi_{\phi}^2\right) \nonumber\\
&=& \rho_{0} \left[\Omega_{\Lambda}+\frac{1+3\ell^2}{4}\frac{(1-\Omega_{\Lambda})}{a^{3(1+\alpha)}}\right]\left[\Omega_{\Lambda} + \frac{(1-\Omega_{\Lambda})}{a^{3(1+\alpha)}}\right]^{-\alpha/(1 + \alpha)}, 
\label{VrhoA}
\end{eqnarray}
where it has been used
\begin{eqnarray}
y^{2}(a) &=& \frac{2}{3}\rho_{\varphi}(a)=\frac{2\rho_{0}}{3} \left[\Omega_{\Lambda} + \frac{(1-\Omega_{\Lambda})}{a^{3(1+\alpha)}}\right]^{1/(1 + \alpha)},\label{aux1}
\end{eqnarray}
and
\begin{eqnarray}
y_\phi^{2}(a) &=&
\rho_{0}\,\frac{(1-\Omega_{\Lambda})}{a^{3(1+\alpha)}}\,\left[\Omega_{\Lambda} + \frac{(1-\Omega_{\Lambda})}{a^{3(1+\alpha)}}\right]^{-\alpha/(1 + \alpha)}.
\label{aux2}
\end{eqnarray}

\begin{figure}
\centering
\includegraphics[width=1.9\columnwidth/2]{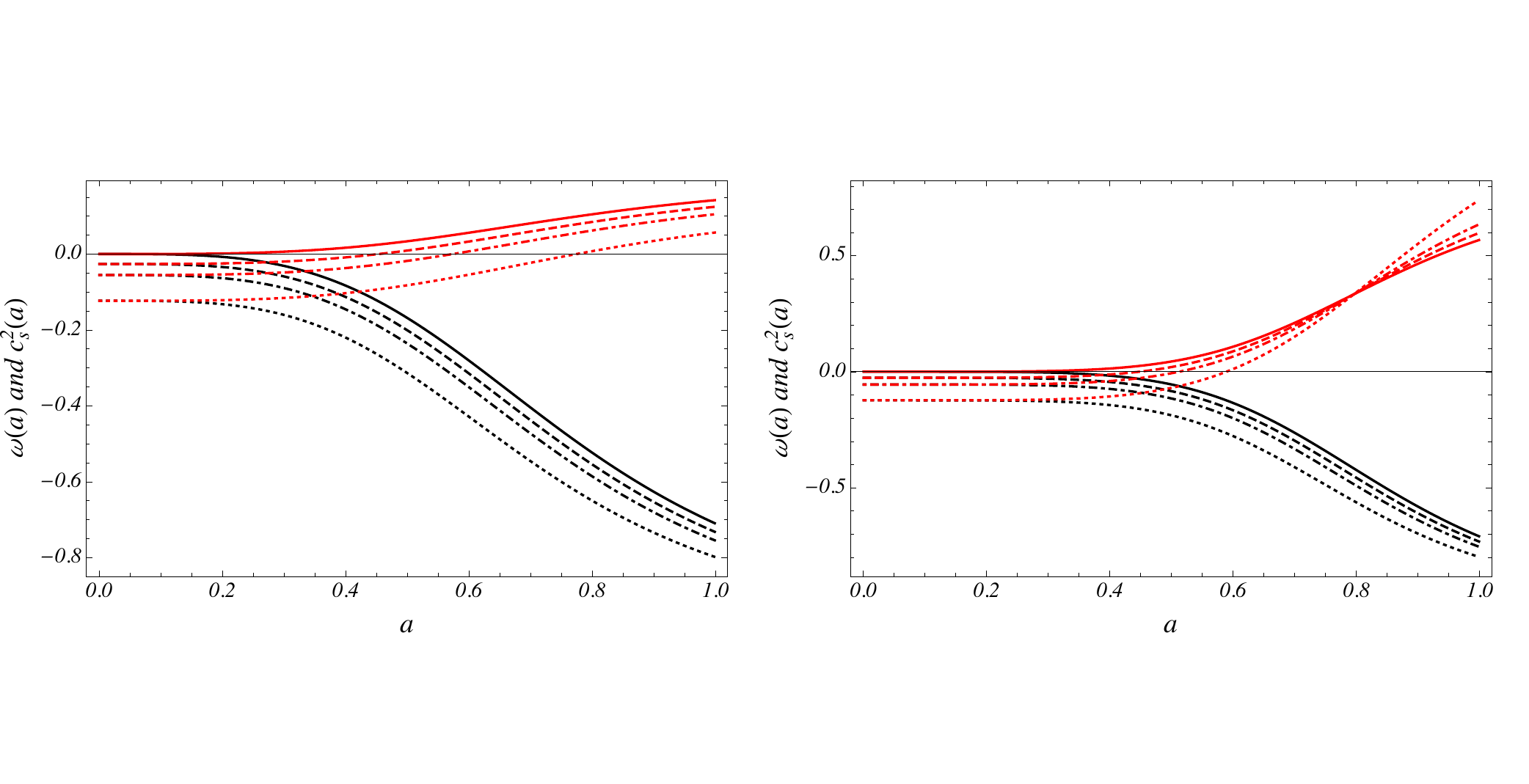}\vspace{-1.8 cm}
\caption{ \footnotesize (Color online) {\em EoS} parameter (black), $w(a)$, and squared speed of sound (red), $c^{2}_{s} = dp/d\rho$, as functions of the scale factor for the cosmological scenario driven by GCG (solid lines) and the $\varphi$-field theories for re-scaled fields with $\varphi_{\phi} = \ell$. Results are for for $\ell = 0.95$ (dashed), $0.90$ (dash-dotted) and $0.80$ (dotted). It has been used $\Omega_{\Lambda} = 0.6825$ from the Planck best fit \cite{planck2020}, and the GCG parameter $\alpha = 0.2$ (first plot) and $0.8$ (second plot).}
\label{FigPapV01A}
\end{figure}

Likewise, for the pressure, $p_{\varphi}(a)$, one has
\begin{eqnarray}
p_{\varphi}(a)
&=& w_{\varphi}^2-\frac{3}{2}w^2(a) \nonumber\\ 
&=& \frac{1}{4} y^2_{\phi}(a)\left(3 + \varphi_{\phi}^2\right) -\frac{3}{2}y^{2}(a)\nonumber\\ 
&=&-\rho_0 \left[\Omega_{\Lambda}+\frac{1-\ell^2}{4}\frac{(1-\Omega_{\Lambda})}{a^{3(1+\alpha)}}\right] \left[\Omega_{\Lambda} + \frac{(1-\Omega_{\Lambda})}{a^{3(1+\alpha)}}\right]^{-\alpha/(1 + \alpha)},
\label{VpA}
\end{eqnarray}
from which the associated equation of state ({\em EoS}) parameter, $w(a) = p(a)/\rho(a)$, and the squared speed of sound, $c^{2}_{s}(a) = dp(a)/d\rho(a)$ can be straightforwardly obtained as depicted in Fig.~\eqref{FigPapV01A}.

\begin{figure}
\centering
\includegraphics[width=1.9\columnwidth/2]{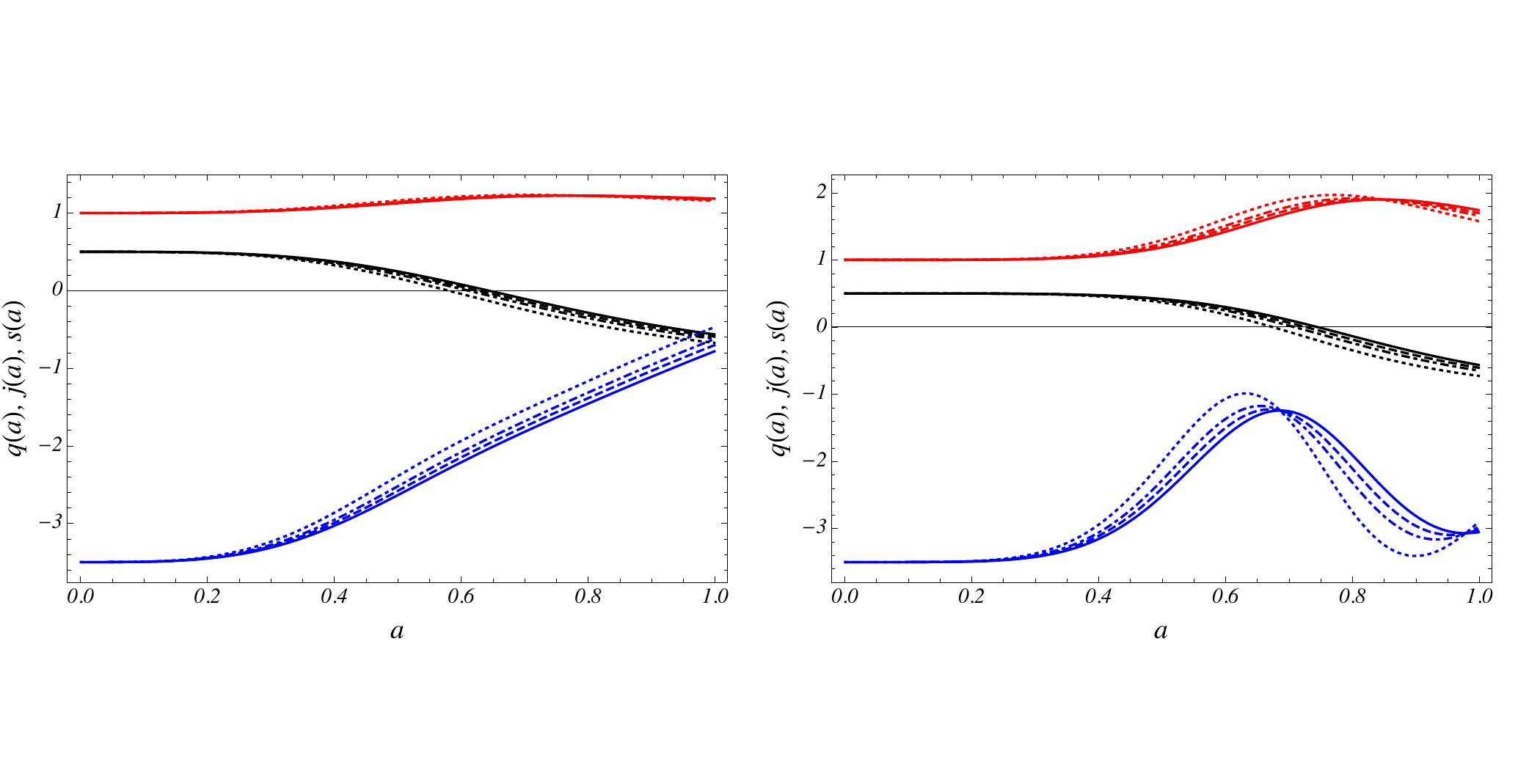}\vspace{-1.8 cm}
\caption{ \footnotesize (Color online) Deceleration (black), $q(a)$, jerk (red), $j(a)$, and snap (blue), $s(a)$, parameters as functions of the scale factor for the cosmological scenario driven by GCG (solid lines) and the $\varphi$-field theories for re-scaled fields with $\varphi_{\phi} = \ell$. Parameter values and line textures are in correspondence with Fig.~\ref{FigPapV01A}.}
\label{FigPapV02A}
\end{figure}

Corresponding deceleration, $q(a)$, jerk, $j(a)$, and snap, $s(a)$, parameters have also been obtained and depicted in Fig.~\eqref{FigPapV02A}. In both figures, results for the $\varphi$-scalar field theory, which encompasses the $c$-field modifications, are compared with the results for the GCG ($\phi$-scalar field theory).
The above results, as it shall be discussed in the following, despite adapting to modified $\Lambda$CDM theories, are problematic in providing conditions for the stable propagation of linear perturbations ($c^{2}_{s} > 0$).
For this reason, a more elaborate solution is proposed.

\subsubsection{Results for a deformed scalar field: $\varphi_\phi = \tanh(\sqrt{3/2}\ell(1+\alpha) \phi)$.}

More suitable results can be obtained from topologically deformed scalar field theories \cite{Ber2013B,Bas01} related by Eq.~\eqref{ee2}. 
That is the case of a kink-like connection between $\varphi$- and $\phi$- theories given by a scalar field deformation 
\begin{equation}
\varphi(\phi) = \sqrt{\frac{2}{3}}\frac{1}{\ell(1+\alpha)}\,\ln\left[\cosh\left(\sqrt{\frac{3}{2}}\ell(1+\alpha)\phi\right)\right],
\label{cfcvA4}
\end{equation}
which, of course, can be replaced by a plethora of deforming kink patterns, $\varphi_\phi$, as the above one.

In this case, energy density and pressure as functions of the scale factor, $a$, can be straightforwardly obtained from Eqs.~\eqref{VrhoA} and \eqref{VpA}, respectively, with $\ell$ replaced by 
\begin{eqnarray}
\varphi_\phi &=& \tanh\left[\sqrt{\frac{3}{2}}\ell (1+\alpha) \phi(a)\right]\nonumber\\ &=& \frac{
\left[\sqrt{\Omega_{\Lambda} + \frac{(1-\Omega_{\Lambda})}{a^{3(1+\alpha)}}}+\sqrt{\frac{(1-\Omega_{\Lambda})}{a^{3(1+\alpha)}}}\right]^{\ell}-\left[\sqrt{\Omega_{\Lambda} + \frac{(1-\Omega_{\Lambda})}{a^{3(1+\alpha)}}}-\sqrt{\frac{(1-\Omega_{\Lambda})}{a^{3(1+\alpha)}}}\right]^{\ell}
}
{
\left[\sqrt{\Omega_{\Lambda} + \frac{(1-\Omega_{\Lambda})}{a^{3(1+\alpha)}}}+\sqrt{\frac{(1-\Omega_{\Lambda})}{a^{3(1+\alpha)}}}\right]^{\ell}+\left[\sqrt{\Omega_{\Lambda} + \frac{(1-\Omega_{\Lambda})}{a^{3(1+\alpha)}}}-\sqrt{\frac{(1-\Omega_{\Lambda})}{a^{3(1+\alpha)}}}\right]^{\ell}
}\nonumber\\
&=& \frac{
\left[
\sqrt{1+\frac{\Omega_{\Lambda}a^{3(1+\alpha)}}{(1-\Omega_{\Lambda})}}+ 1
\right]^{\ell}
-
\left[
\sqrt{1+\frac{\Omega_{\Lambda} a^{3(1+\alpha)}}{(1-\Omega_{\Lambda})}}-1
\right]^{\ell}
}
{
\left[
\sqrt{1+\frac{\Omega_{\Lambda}a^{3(1+\alpha)}}{(1-\Omega_{\Lambda})}}+ 1
\right]^{\ell}
+
\left[
\sqrt{1+\frac{\Omega_{\Lambda} a^{3(1+\alpha)}}{(1-\Omega_{\Lambda})}}-1
\right]^{\ell}
}.
\label{VrhoB}
\end{eqnarray}

\begin{figure}
\centering
\includegraphics[width=1.9\columnwidth/2]{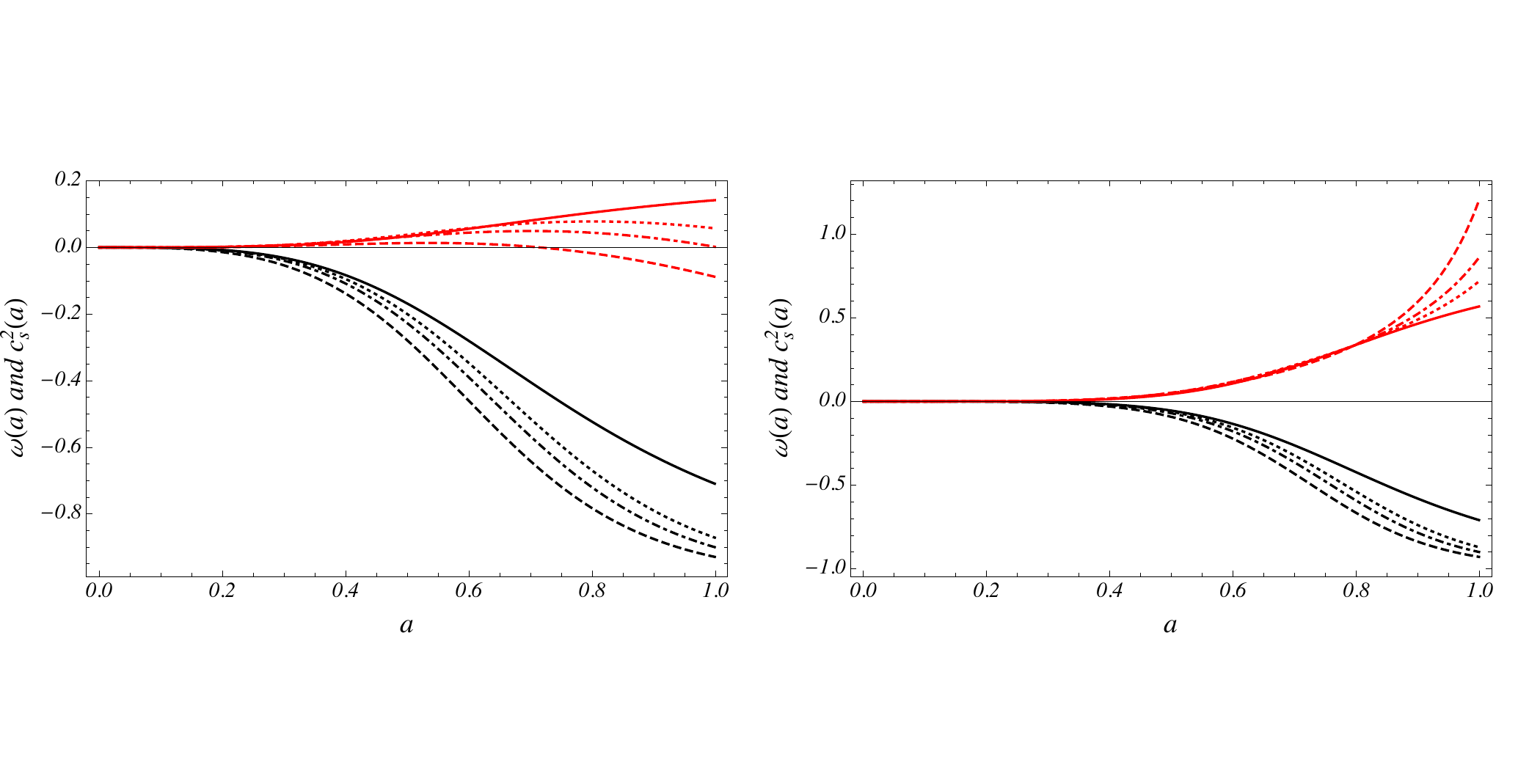}\vspace{-1.8 cm}
\caption{ \footnotesize (Color online) {\em EoS} parameter (black), $w(a)$, and squared speed of sound (red), $c^{2}_{s} = dp/d\rho$, as functions of the scale factor for the cosmological scenario driven by GCG (solid lines) and the $\varphi$-field theories for deformed fields with $\varphi_{\phi} = \tanh(\sqrt{3/2}\ell(1+\alpha) \phi)$. Results are for for $\ell = 0.8$ (dashed), $1.0$ (dash-dotted) and $1.2$ (dotted). It has been used $\Omega_{\Lambda} = 0.6825$ from the Planck best fit \cite{planck2020}, and the GCG parameter $\alpha = 0.2$ (first plot) and $0.8$ (second plot).}
\label{FigPapV01B}
\end{figure}

\begin{figure}
\centering
\includegraphics[width=1.9\columnwidth/2]{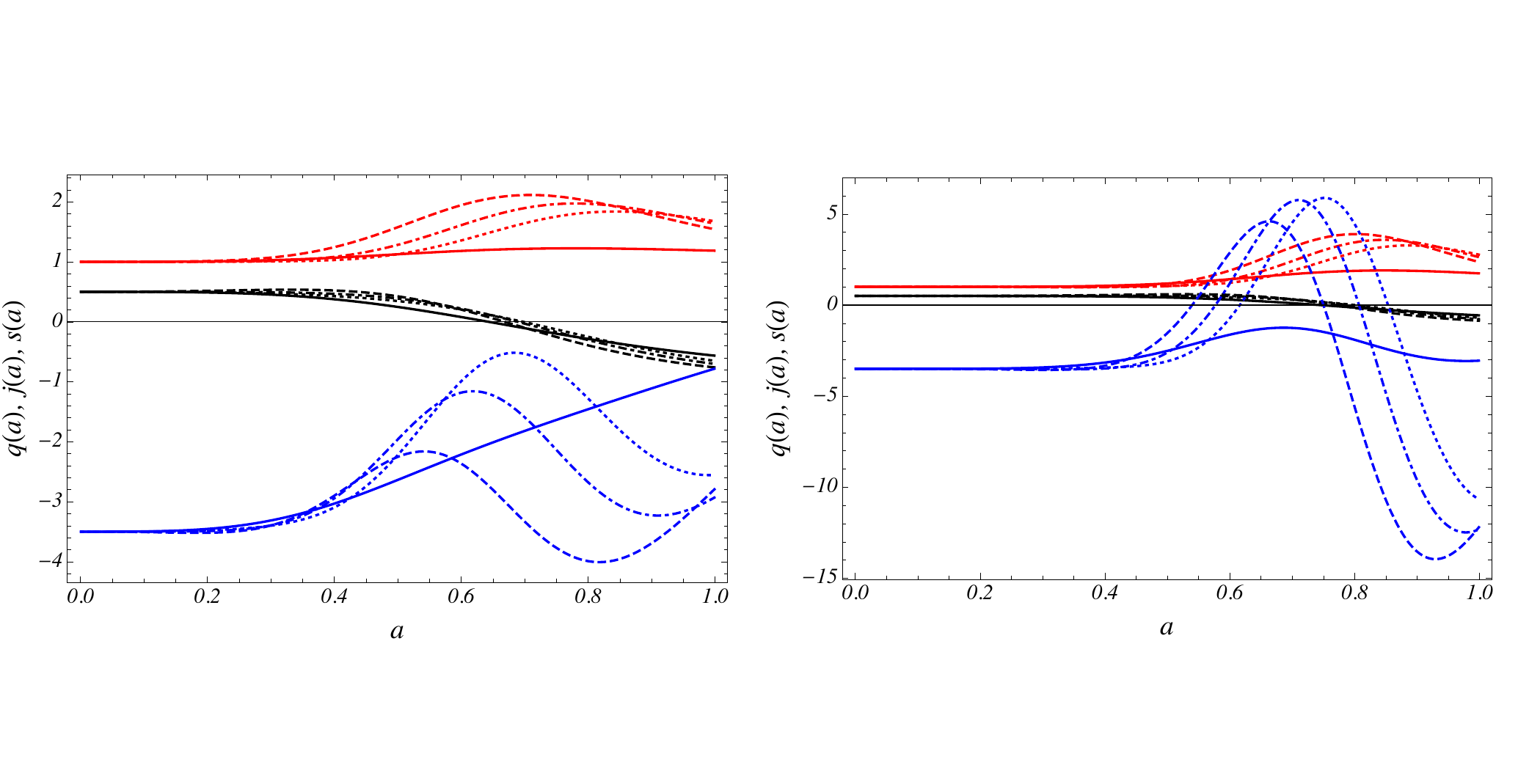}\vspace{-1.8 cm}
\caption{ \footnotesize (Color online) Deceleration (black), $q(a)$, jerk (red), $j(a)$, and snap (blue), $s(a)$, parameters as functions of the scale factor for the cosmological scenario driven by GCG (solid lines) and the $\varphi$-field theories for deformed fields with $\varphi_{\phi} = \tanh(\sqrt{3/2}\ell(1+\alpha) \phi)$. Parameter values and line textures are in correspondence with Fig.~\ref{FigPapV01B}.}
\label{FigPapV02B}
\end{figure}

The {\em EoS} parameter, $w(a) = p/\rho$, and the squared speed of sound, $c^{2}_{s}(a) = dp/d\rho$ can be computed, and are depicted in Fig.~\eqref{FigPapV01B}.
Similarly, deceleration, $q(a)$, jerk, $j(a)$, and snap, $s(a)$, parameters are also obtained and depicted in Fig.~\eqref{FigPapV02B}.

The most relevant feature identified in Fig.~\ref{FigPapV01B} concerns the stability of the propagation of linear perturbations, that is $c^2_s > 0$, for the entire universe expansion phase.

Results for $s(a)$ indicates that the perturbative formulation of the problem is not well posed, as one can infer through the higher order perturbations. At early times, contributions from increasing order parameters, namely for $j(a)$ and $s(a)$, suggest that perturbative modifications to the dark sector background (black lines) are problematic and favors the inclusion of $c$-field contributions in the context of a non-perturbatively modified scalar field theory as proposed here.

\subsubsection{The contribution to the Hubble expansion rate}

From the previous results, one notices that, for deformed scalar fields, and even for re-scaled fields, the early time dark sector contributions remains unaffected, with cosmological changes being manifest at late times.
This suggests the modified scalar field theories parameterized by $c$-fields as potential drivers of Hubble expansion rate modifications at late times.

The $c$-field contribution to the square of $H$ (cf. Eq.~\eqref{ee2mBB}), is given by
\begin{equation}
- \frac{\dot{c}^2}{3} = -\frac{1}{2}\rho_{0}(1-\ell^2) \,\frac{(1-\Omega_{\Lambda})}{a^{3(1+\alpha)}}\,\left[\Omega_{\Lambda} + \frac{(1-\Omega_{\Lambda})}{a^{3(1+\alpha)}}\right]^{-\alpha/(1 + \alpha)},
\label{ee2mBB01}
\end{equation}
which evolves as a subtracting matter contribution at early times ($a \ll 1$), with
\begin{equation}
- \frac{\dot{c}^2}{3} \approx -\frac{1}{2}\rho_{0}(1-\ell^2) (1-\Omega_{\Lambda})^{1/(1 + \alpha)}{a^{-3}}
\label{ee2mBB02}
\end{equation}
being less dominating at late times, with
\begin{equation}
- \frac{\dot{c}^2}{3} \propto - {a^{-3(1 + \alpha)}}.
\label{ee2mBB03}
\end{equation}

\begin{figure}
\centering
\includegraphics[width=1.9\columnwidth/2]{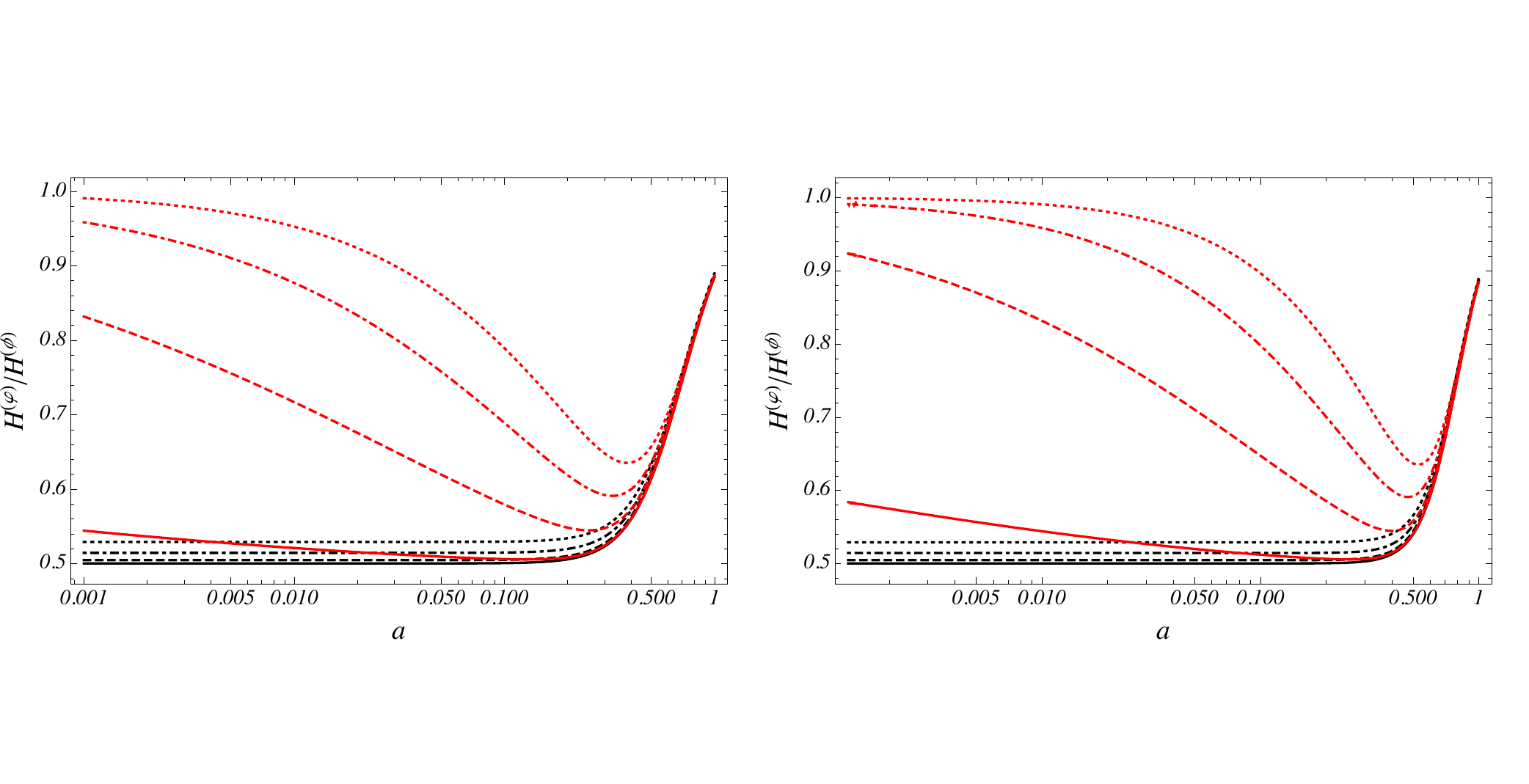}\vspace{-1.8 cm}
\caption{ \footnotesize (Color online) Relative Hubble expansion rate, $\frac{H_{(\varphi)}}{H_{(\phi)}}$ as functions of the scale factor for the cosmological scenario driven by $\varphi$-field theories for re-scaled fields (black) with $\varphi_{\phi} = \ell$, and deformed fields (red) with $\varphi_\phi = \tanh(\sqrt{3/2}\ell(1+\alpha) \phi)$. Results are for for $\ell = 0.02$ (solid), $0.08$ (dashed), $0.14$ (dash-dotted) and $0.2$ (dotted). Again, $\Omega_{\Lambda} = 0.6825$ and $\alpha = 0.2$ (first plot) and $0.8$ (second plot).}
\label{FigRate}
\end{figure}

From Eq.~\eqref{ee2mcc}, the impact of the $c$-field contributions mapped by the $\varphi$-scalar field theory on the Hubble expansion rate can be evaluated by
\begin{equation}
\frac{H_{(\varphi)}}{H_{(\phi)}}= \sqrt{1 - \frac{1}{2}\frac{y^2_{\phi}}{y^2}\left(1 - \varphi_{\phi}^2\right)},
\label{ee2mdd}
\end{equation}
In the case of the GCG, for which $\phi(a)$ decreases with $a$, the above choices from Eqs.~\eqref{cfcvA3} and \eqref{cfcvA4} leads to modified Hubble expansion rates, where $H_{(\varphi)}$ replaces $H_{(\phi)}$ as identified in Fig.~\ref{FigRate}.
From Fig.~\ref{FigRate}, one notice that both approaches, for $\varphi_{\phi} = \ell$ and $\varphi_\phi = \tanh(\sqrt{3/2}\ell(1+\alpha) \phi)$, can be phenomenologically adjusted to fit the evolution of $H$ from early to late times. Of course, as it was emphasized, $\varphi_{\phi}$ is a free degree of freedom just constrained by the inclusion of the $c$-field DoF in the formalism.

In summary, such a dynamical behavior can be attributed to the scalar field associated with the GCG when coupled with a $c$-field DoF, allowing the usual phenomenological constraints of the GCG to be relaxed. As a result, a broader range of cosmological scenarios can be reconciled with effective GCG models.
A smooth change in the Universe's equation of state is allowed under two equivalent scenarios: either $c$-field contributions are included and the original scalar field derived from the GCG is maintained, or an exact solution to the Friedmann equation for a mapped $\varphi$-field theory is admitted.

\section{Conclusions}

The HN creation field cosmology has been investigated, and its correspondence with scalar field models for the dark sector, has been established. Within a Hamilton-Jacobi reconstruction framework, it was demonstrated that the $c$-field can be effectively mapped onto an equivalent scalar field theory replicating GCG-like behavior, thereby allowing a putative unification of dark matter and dark energy into a single dynamical component. 

The resulting {\em EoS} and the background for the evolution of linear perturbations have been examined for re-scaled and deformed scalar field theories. For the latter ones, the propagation was found to remain stable throughout the cosmic expansion, as indicated by a strictly positive sound speed, $c_s^2 > 0$. Different functional forms of the auxiliary scalar coupling, such as $\varphi_{\phi} = \ell$ and $\varphi_{\phi} = \tanh(\sqrt{3/2}\ell(1+\alpha)\phi)$, were considered as capable of capturing the evolution of the Hubble expansion rate across cosmological epochs. As $\varphi_{\phi}$ is treated as a free function constrained only by the dynamical structure of the $c$-field, a broader class of solutions becomes accessible.

The coupling of the $c$-field to the GCG scalar field introduces a dynamical mechanism that allows for relaxing conventional GCG constraints. This enables a smooth evolution of the Universe's equation of state without altering the underlying scalar field formulation. An exact solution to the Friedmann equation emerges within an alternative $\varphi$-field theory, further reinforcing the model's internal consistency.

The $c$-field cosmology, when reinterpreted within a scalar field formalism, presents a stable, cosmologically consistent, and theoretically unified alternative to the standard $\Lambda$CDM model. The framework accommodates both late-time acceleration and early-universe dynamics while offering flexibility to address existing cosmological tensions \cite{cosmoverse,pheno,Riess24,Var24,Ber25}, such as those related to the Hubble constant, $H_0$.
Of course, in this case, a deeper phenomenological analysis involving different datasets of early-time CMB measurements \cite{planck2020} and late-time Universe surveys \cite{Var24,Pantheon} must be considered.

In conclusion, our analysis provides an alternative framework for understanding the interplay between dark matter and dark energy, highlighting the potential for further research into unified cosmological models that may offer novel insights into the fundamental constituents of the Universe.

{\em Acknowledgments --- The work of AEB is supported by the Brazilian Agency CNPq (Grant No. 301485/2022-4, National Council for Scientific and Technological Development --- CNPq).}

\end{document}